%% file: main.tex
\documentclass[lettersize,journal]{IEEEtran}
\usepackage{amsmath,amsfonts}
\usepackage{algorithmic}
\usepackage{algorithm}
\usepackage{array}
\usepackage[caption=false,font=normalsize,labelfont=sf,textfont=sf]{subfig}
\usepackage{textcomp}
\usepackage{stfloats}
\usepackage{url}
\usepackage{verbatim}
\usepackage{graphicx}
\usepackage{cite}
\usepackage{pgfplots}
\usepackage{amsmath}
\usepackage[numbers,sort&compress]{natbib}
\usepackage[inkscapeformat=png]{svg}
\usepackage{caption}
\begin{document}
\pgfplotsset{compat=1.18} 

\title{Quasiparticle Dynamics in NbN Superconducting Microwave Resonators at Single Photon Regime}

\author{Paniz Foshat, Shima Poorgholam-khanjari, Valentino Seferai, Hua Feng, Susan Johny, Oleg A. Mukhanov, Matthew Hutchings, Robert H. Hadfield, Martin Weides, and Kaveh Delfanazari

        % <-this % stops a space

\thanks{Dated June 21, 2025}
\thanks{ This study made use of the University of Glasgow James Watt Nanofabrication Centre (JWNC), and we thank JWNC technical staff for their support.
}% <-this % stops a space
\thanks{Paniz Foshat, Shima Poorgholam-Khanjari, Valentino Seferai, Hua Feng, Susan Johny, Robert H. Hadfield, Martin Weides, and Kaveh Delfanazari are with the Electronics and Nanoscale Engineering Division, James Watt School of Engineering, University of Glasgow, Glasgow, United Kingdom. (Corresponding author: kaveh.delfanazari@glasgow.ac.uk)}
\thanks{ Oleg A. Mukhanov is with SEEQC, Elmsford, NY 10523 USA. Matthew Hutchings is with SeeQC, London, United Kingdom.}
}

% The paper headers
\markboth{}
{Shell \MakeLowercase{\textit{et al.}}: A Sample Article Using IEEEtran.cls for IEEE Journals}

%\IEEEpubid{0000--0000/00\$00.00~\copyright~2021 IEEE}
% Remember, if you use this you must call \IEEEpubidadjcol in the second
% column for its text to clear the IEEEpubid mark.

\maketitle

\begin{abstract}
Exchanging energy below the superconducting gap introduces quasiparticle energy distributions in superconducting quantum circuits, which will be responsible for their decoherence. This study examines the impact of quasiparticle energy on the performance of  NbN superconducting microwave coplanar waveguide resonators on silicon chips.
We measured the resonance frequency and internal quality factor in response to temperature sweeps to evaluate the effect of quasiparticle dynamics. Moreover, by calculating the complex conductivity of the NbN film, we identified the contribution of quasiparticle density to the experimental results.  
\end{abstract}


\section{Introduction}
\IEEEPARstart{B}{oosting} coherence time has become crucial for maximizing the efficiency of superconducting quantum circuits \cite{siddiqi2021engineering}, especially in Cooper pair transistors \cite{aumentado2004nonequilibrium}, kinetic inductance detectors \cite{zmuidzinas2012superconducting,day2003broadband,morozov2021superconducting}, and superconducting qubits \cite{lenander2011measurement}.

\begin{figure}[t]
\centering
\subfloat[]{\includegraphics[scale=0.15]{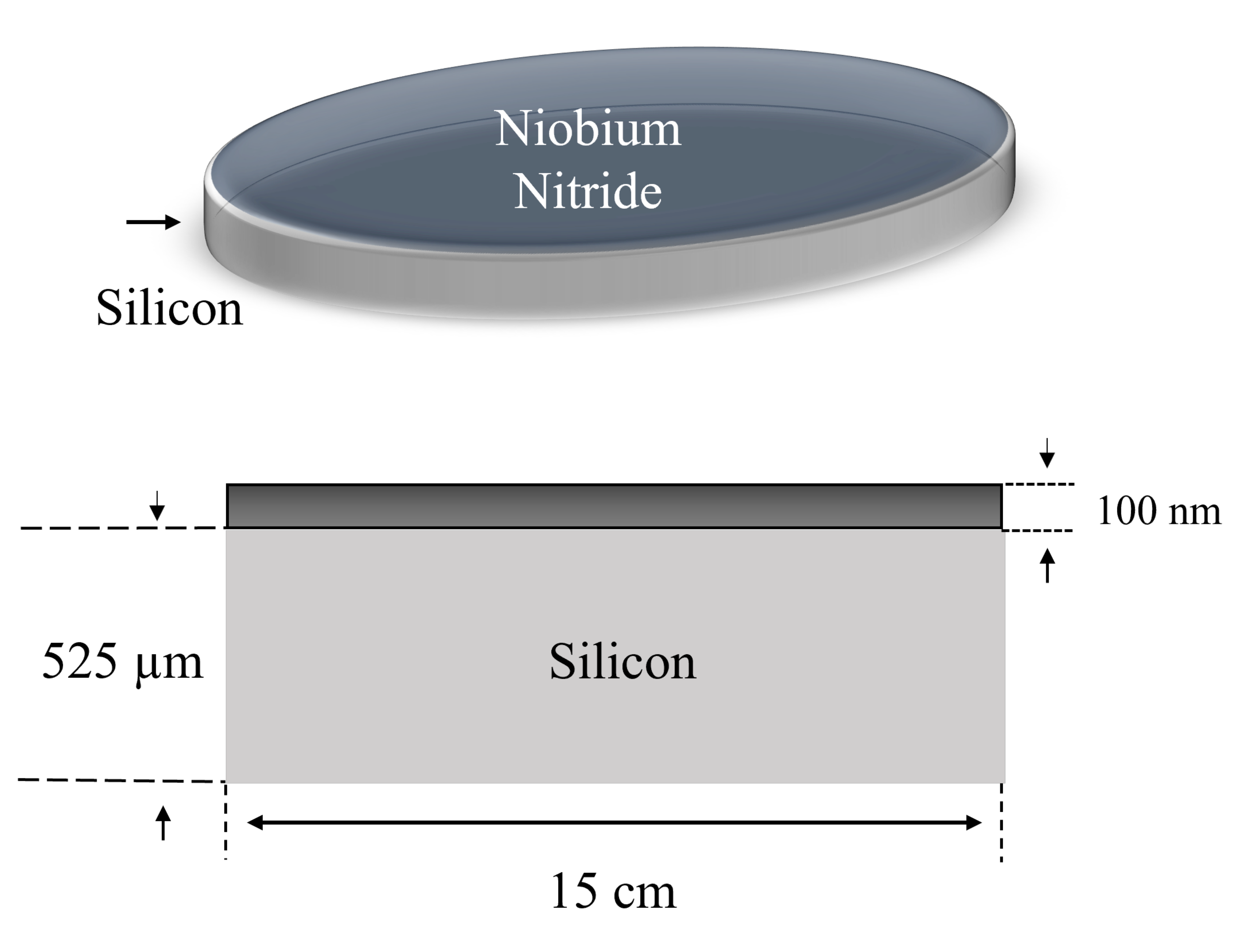}
\label{fig_1_a}}
\hfill
\centering
 \subfloat[] {\includegraphics[]{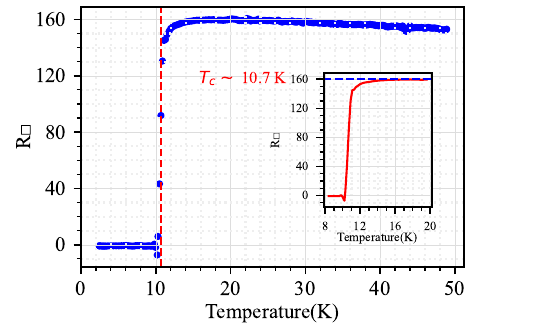}
 \label{fig1_b}}
 \caption{ (a) Top view and cross-section illustrations of a 100 nm NbN thin film sputtered on a 525 $\mu$m thick silicon wafer for cryogenic DC transport and RF spectroscopy measurements.
(b) Sheet resistance versus temperature for the NbN/silicon chip. RRR is estimated to be $\frac{R_{300 K}}{R_{T_c}} \approx 0.98 $.}
 \label{fig_1}
 \end{figure}

The presence of charge carriers that participate in superconducting condensate, known as quasiparticle poisoning \cite{aumentado2023quasiparticle}, is currently one of the primary reasons limiting the performance of superconducting quantum circuits\cite{burnett2018noise,wisbey2010effect,pitsun2020cross,wang2014measurement}. Moreover, recent studies indicate that in multi-qubit systems, quasiparticle emergence can suppress the qubit relaxation time since quasiparticles are coupled with all qubits in a wafer-scale chip \cite{wilen2021correlated}.
Research demonstrates that quasiparticles' impact on the performance of quantum circuits can be reduced or eliminated by carefully selecting materials \cite{catelani2022using} and engineering designs \cite{pan2022engineering,riwar2019efficient,kamenov2023suppression}. For example, including quasiparticle' traps made of a lower gap superconductor on the quantum circuit is effective for this purpose \cite{henriques2019phonon}.
Therefore, characterizing the quasiparticle dynamics of superconductor films is important in the development of superconducting qubits, and quantum circuits \cite{wilen2021correlated,connolly2023coexistence}. \par
Niobium Nitride (NbN) is a type II superconductor with promising applications in unconventional superconducting quantum circuits \cite{gustavsson2016suppressing}. Recent studies have shown that superconductors-semiconductor nanodevices can host Majorana quasiparticles or be utilized in Andreev-based superconducting circuits \cite{aguado2020perspective,prada2020andreev}. Given that NbN can operate in quantum circuits at higher temperatures, wider frequencies and under higher magnetic fields, it is essential to investigate the behaviour and presence of quasiparticles across different temperatures. To do so, we fabricate NbN superconducting microwave coplanar waveguide resonators (CPW), as they are fundamental elements in most quantum circuits. \par
For this study, we selected a 100 nm thickness ($t$) NbN film as a trade-off between low-loss characteristics and robustness against magnetic fields of the superconducting film. While thicker films are typically used in transmon qubits, they are often more susceptible to the extensive magnetic field effects on CPWs' properties, such as resonance frequency ($f_r$), and internal quality factor ($Q_i$), due to the emergence of Abrikosov vortices. Our previous work demonstrated that 100 nm NbN films provide a high $Q_i$ with and without the presence of the magnetic field \cite{foshat2025characterizing}, making them a suitable choice for investigating quasiparticle dynamics in hybrid superconductor-semiconductor circuits.
Moreover, CPWs are a powerful and valuable tool for measuring the losses of superconducting films \cite{alexander2022power}.
CPW resonators characterize losses such as two-level system (TLS) defects \cite{pappas2011two}, non-equilibrium quasiparticles  \cite{grunhaupt2018loss,noguchi2021contribution,barends2007niobium}, vortices \cite{song2011microwave,muller2022magnetic}, and microwave radiation \cite{sage2011study,hahnle2020suppression} in superconducting circuits. 
Calculating and modelling TLS defects is relatively straightforward in superconducting CPW; however, resolving quasiparticle losses requires intensive computational calculations. Theoretical predictions indicate that Cooper pair breaking should be negligible at temperatures much lower than the critical temperature ($T_c$) due to insufficient thermal energy to break the pairs. However, experimental results in materials such as aluminium (Al) \cite{de2012microwave,de2021strong} and titanium nitride (TiN) \cite{hu2024investigation} show a persistent presence of quasiparticles even at temperatures significantly below $T_c$.
While limited research has been done on quasiparticle density in Nb-based superconducting films \cite{zhang2020quasiparticle, drimmer2024effect}, there is a significant gap in the research on quasiparticle density focusing on Nb-based CPW.

\par Here, we will have a more in-depth discussion on characterizing quasiparticles' emergence in superconducting microwave CPW resonators.
We model TLS loss with a standard conventional model \cite{pappas2011two,foshat2025characterizing,poorgholam2024engineering}. To describe quasiparticle loss, we calculate the complex conductivity of the NbN film based on Mattis-Bardeen theory \cite{barends2009photon,mattis1958theory}. 

\par On the measurement side, we conduct a DC measurement of the NbN film to find $T_c$ and resistivity in the normal state ($R_\square$). Moreover, we perform cryogenic microwave spectroscopy of the NbN superconducting CPW resonators to find the effect of temperature sweep $T < \frac{T_{c}}{3}$ on the measured $Q_{i}$ and resonance frequency shift ($\Delta f$). We successfully maintain a high $Q_{i} > 7\times10^3$ at temperatures below $3$ K. We prove the reliability of the $Q_i$ measurements by comparing the measurement results with analytical calculations.
 \begin{figure}[t]
 \centering
\hspace{-0.2cm}\subfloat[] {\includegraphics{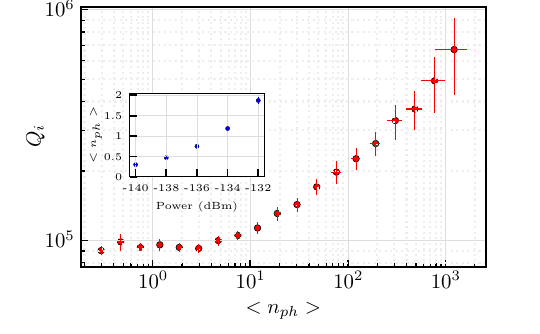}}
\vfill
\subfloat[]{\includegraphics{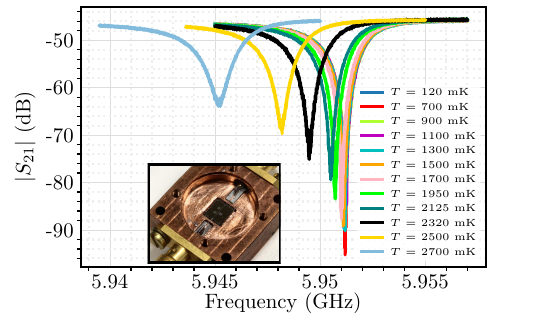}}
 % \includegraphics[scale=0.2]{Figures/Tempsweep.png}
%\includesvg [width = 3 in]{Tempsweep_latexcompatible_1e.svg}
    %\input{samplebox.pgf}
 \caption{ (a) $Q_{i}$ in the photon number ranges from $<n_{ph}>$ = $10^{-1}$ to $10^3$ corresponding to the power ranges from $P_{in}$ = -140 dBm to -100 dBm at $T$ = 26 mK, and $f_r$ = 5.95 GHz, adapted from \cite{foshat2025characterizing}. The inset plot shows the approximate $<n_{ph}>$ concerning input power inside the transmission line in the CPWs. In this plot, the input power to reach the single photon regime is $P_{in}$ $\simeq$ -135 dBm. The vertical lines show $Q_i$ uncertainty range, and the horizontal lines show $< n_{ph} >$ uncertainty range. The details about photon number calculation are illustrated in the appendix. (b) Microwave spectroscopy of NbN CPWs resonator at 0.1 K $< T <$ 3 K (in single photon regime) at $P_{in}$ $\simeq$ -135 dBm. The inset shows the optical image of the NbN CPW chip being packaged for cryogenic microwave spectroscopy.}
 \label{fig_2}
 \end{figure}
\begin{table}[h]
 \caption{ NbN film sputtering process information.}
\begin{center}
 \begin{tabular}{||c c c c||} 
 \hline
 \textbf{Ar} & \textbf{N$_2$} & \textbf{Base current} & \textbf{Sputtering time} \\ [0.5ex] 
 \hline\hline
 25 sccm & 3 sccm & 0.85 A & 15 min \\
 \hline
\end{tabular}
\end{center}
  \label{tab:table2}
\end{table}
\section{Fabrication method}
In our design, three resonators are capacitively coupled to a common feedline, with 4 $\mu$m width and 2 $\mu$m gap. \par
To fabricate the superconducting circuit, we dipped a high-resistivity silicon wafer into Buffer Oxide Etchant (BOE) to strip the SiO$_x$ layer and prepare it for superconducting film deposition. Immediately after, the wafer was placed in a sputtering system to form a $100$ nm NbN film. The NbN film was produced through reactive DC sputtering using an argon (Ar) and nitrogen (N$_2$) plasma \cite{banerjee2018optical}. Table~\ref{tab:table2} illustrates the details of the sputtering process of the NbN film. 
Two samples of the chip were used after dicing: one to pattern the resonators with standard e beam lithography, followed by CF$_4$ anisotropic dry etch, and the other was used for DC measurement.\par

It should be noted that silicon was selected as the substrate in our CPWs chips because of its compatibility with the fabrication process of advanced semiconductor devices and hybrid superconductor–semiconductor circuits. Although other substrates, such as sapphire, offer advantages in terms of reduced dielectric loss, silicon remains the preferred choice because of its compatibility with industrial semiconductor fabrication processes. However, coherence times in these types of hybrid quantum circuit remain a key limitation to further scaling them. Addressing this challenge requires comprehensive research, and we have taken the initial step towards it in this paper. As for the superconducting material, we concentrated on depositing NbN films to achieve low-loss superconducting films that are well-suited for the simple fabrication of robust hybrid superconductor–semiconductor circuits, such as through room temperature deposition of the NbN film. \par
\begin{figure}[!t]
\centering
\subfloat[]{\includegraphics[]{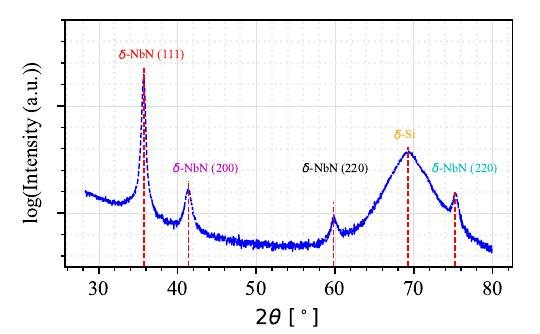}
 \label{XRD1}}
 \centering
 \vfill
 \subfloat[]{\includegraphics[]{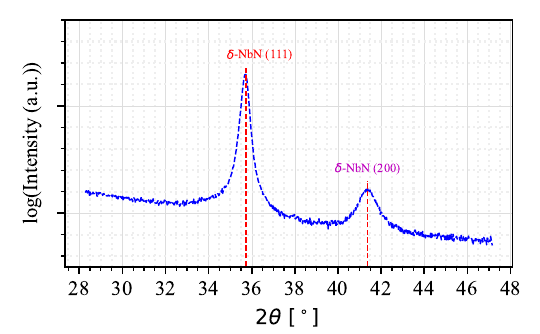}
 \label{XRD}}
 \caption{ (a) The XRD pattern of the NbN film. (b) Zoomed-in region (30°–48°) showing diffraction peaks associated with the superconducting phases of NbN.}
 \label{fig_XRD}
 \end{figure}

To further investigate the structure and composition of our NbN film, we examined our sample with X-ray diffractometry to determine its orientation and phase purity. Fig.~\ref{fig_XRD} (a) displays the X-ray diffraction (XRD) pattern of our NbN film. In Fig.~\ref{fig_XRD} (b), we observed the prominent peaks corresponding to the (111) and (200) oriented grains of cubic $\delta$-NbN film at 35.7$^o$  and 41.3$^o$, respectively. We also noticed two further peaks at 59.8$^o$ and 75.2$^o$, which can be assigned to the grains of (022) and (222) orientation of NbN film \cite{sowa2017plasma,knehr2021wafer,shi2022nbn}. The full width at half maximum (FWHM) of the (111) and (200) peaks were measured to be 0.47$^o$ and 1.36$^o$, respectively. The sharper (111) peak is useful, as it is often associated with improved electrical and superconducting properties in $\delta$-NbN films \cite{sowa2017plasma}. Peak broadening, as observed in the (200), (022), and (222) reflections, may be attributed to reduced crystallite size, microstrain, or structural defects. The calculated lattice constant ($a$) from the (111), (200), (022), and (222) peaks range from 4.35 $\text{\AA}$ to 4.37 $\text{\AA}$, lower than the ideal $\delta$-NbN value of a = 4.45 $\text{\AA}$. This slight difference indicates a non-ideal lattice possibly due to strain, mechanical stress and induced defects in NbN film \cite{lennon2023high,alfonso2010influence}. Table~\ref{XRD} summarizes the above information.
\begin{table}[h]
 \caption{ XRD pattern data including: orientation, peak position, FWHM, grain and lattice size of NbN film.}
\begin{center} \hspace{-0.8 cm}
\scalebox{1.1}{
 \begin{tabular}{||c c c c c||}
 \hline
\textbf{Orientation} &  \textbf{2$\theta$ ($^o$)} &  \textbf{FWHM ($^o$)} &  \textbf{Grain Size (nm)} &  \textbf{$a$  ($\text{\AA}$}) \\ [0.5ex] 
 \hline\hline
 (111) &  35.73 & 0.47 & 176.07 & 4.35 \\
(200) &   41.38 & 1.36 & 62.54 & 4.36 \\
(022)  & 59.81 & 1.67 & 54.89 &  4.37 \\
(222) &  75.23 & 1.44 & 69.57 & 4.37 \\
 \hline
\end{tabular}}
\end{center}
  \label{XRD}
\end{table}
\section{DC measurement}
 We characterized the NbN film by performing cryogenic DC transport measurements of a 15 mm $\times$ 15 mm chip between the temperature ranges from 2 K to room temperature to find the film parameters, such as $T_{c}$, sheet resistance $R_\square (T_c)$ and residual resistive ratio (RRR). Fig.1(a) shows the schematic illustration of the wafer and the detail of the diced silicon wafer after dicing (bottom) used for DC measurements. Fig.1(b) illustrates the extracted $T_{c} = 10.7 \pm 0.5$ K and $R_\square (T_c)= 159.5 \: \Omega $, respectively. Although achieving NbN films with higher $T_c$ is possible \cite{zhang2019superconductivity,hazra2016superconducting}, through varying deposition parameters such as substrate heating, the room temperature magnetron sputtering employed in this study was selected to ensure compatibility with the fabrication process of hybrid superconductor–semiconductor quantum circuits. However, this process remains open to further optimization, such as by modifying sputtering parameters and applying post-deposition treatments, to improve superconducting $T_c$ while preserving compatibility with the superconductor–semiconductor fabrication process. We note that the 
$T_c$ reported here is not the upper limit; higher values ($T_c \approx$ 12.5 K) were later obtained on different substrates with improved sputtering tool conditions, though these results are beyond the scope of this work.
 
\section{Temperature Dependent Microwave Spectroscopy of CPW Resonators}
Theoretically, any phonon or photon with energy greater than the superconducting energy gap (2$\Delta$) can break Cooper pairs and generate quasiparticles. These quasiparticles will create resistive and inductive channels in the CPW resonators
 \cite{alexander2021measuring}. Emerging new inductive and resistive channels based on quasiparticle density in CPW will affect its propagation properties. Calculating the superconducting resonator's surface impedance will help predict and confirm $f_r$ and $Q_i$ due to variations in quasiparticle density. The surface impedance of the superconductor depends on the film thickness, $d$, mean free path, $l$, coherence length, $\xi$, and penetration length, $\lambda$. In the superconducting dirty limit condition, where $\l \ll \xi$ and $\l \ll \lambda$, surface impedance and quasiparticle loss can be written as \cite{gao2008physics,alexander2021measuring}: 
\begin{equation}\label{equation1}
    Z_s(T) = \sqrt{ \frac{j\mu_0\omega}{\sigma_1(T)-j\sigma_2(T)} }= R_s + j \omega L_s
\end{equation}

 \begin{equation}\label{equation2}
    \delta_{qp, Theory}(T) \sim \frac{Re (Z_s(T))}{\omega \: (Im(\frac{Z_s(T)}{\omega}))+L_g)} 
\end{equation}

Where $\sigma_1$ is the real part of the complex conductivity, $\sigma_2$ is the imaginary part of the complex conductivity, $L_g$ is the geometrical inductance \cite{goppl2008coplanar}, $\mu_0$ is the free space permeability, $\omega$ is the radian frequency, and  $\delta_{qp, theory}$ is theoretical quasiparticle loss based on the CPW's complex conductivity.  Therefore, by calculating the superconductive film's complex conductivity, we can find the contribution of quasiparticle loss in superconductive CPW resonators. To evaluate the effect of resistive and inductive channels, we perform microwave spectroscopy measurements of our $d$ = 100 nm thick NbN resonators in single photon regime at sub-Kelvin temperature ranges and calculate superconducting film complex conductivity, which will be discussed in more detail in the next section.
\par
\subsection{Cryogenic Measurement of Microwave CPW Resonators}
For microwave spectroscopy measurements, the CPW resonator chip was glued into a copper sample box and wirebound with aluminium wire. The sample box was mounted inside an adiabatic demagnetization refrigerator (ADR). Note that the temperature sensor of the millikelvin stage was RuO$_x$, an effective and accurate low-temperature thermometer. In the ADR setup, complex microwave transmission spectroscopy $S_{21}$ was conducted using an MS4642B Keysight Vector Network Analyzer (VNA). To achieve the single photon regime, the input power of the VNA ($P_{VNA}$) was set to -25 dBm. Subsequently, the VNA's input line was significantly attenuated by -50 dB at room temperature. Additionally, the VNA signal was attenuated by -20 dB at each stage of the ADR (at temperatures of 50 K, 4 K, and 0.8 K), resulting in a total attenuation ($P_{att}$) of -110 dB in the measurement setup. As this total attenuation was applied before the signals reached the sample's feed line, it effectively suppressed thermal noise, enabling the resonator to operate in the single-photon regime.
Note that according to Fig.2(a), which references the data from our previous measurement \cite{foshat2025characterizing}, the power range necessary to achieve the single photon regime in the CPW feedline is between -134 dBm and -136 dBm. This precisely matches the power level selected in this measurement setup. For all subsequent experiments, we consistently use an input power of $P_{in} = P_{VNA} + P_{att}$ = -135 dBm. After the input signal went through the sample, an isolator was placed at the 4 K stage to prevent transmission signals from being contaminated by spurious reflections from the sample box. The signal was then amplified by a high electron mobility transistor (HEMT) at the 4 K stage, providing 40 dB amplification. Additionally, a room temperature amplifier added an extra 45 dB amplification.\par
To investigate the temperature effects on propagation properties of NbN-based microwave CPWs, we measured full complex scattering data of our chips, Fig.2(b), within the temperature ranges of 0.1 K $< T < $ 3 K. We observed noticeable $\Delta f$ and amplitude reduction in $|S_{21}|$. 
\begin{figure}[!t]
  \hspace{-0.5 cm}
\subfloat[]{\includegraphics[]{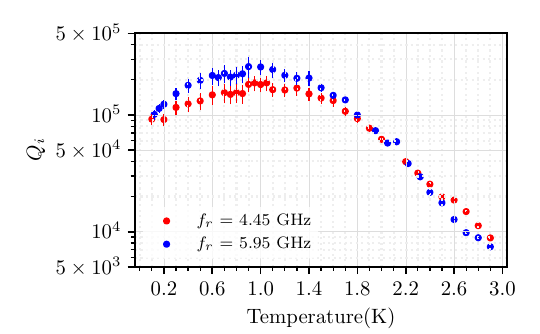}
 \label{fig3_a}}
\vfill
 \centering
  \hspace{-0.5 cm}
 \subfloat[] {\includegraphics[]{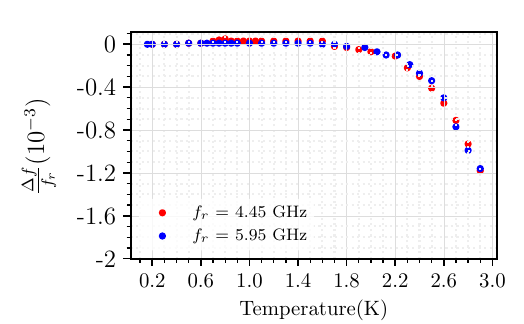}
 \label{fig3_b}}
 \caption{(a) $Q_i$ at single photon regime versus temperature for $f_r$ = 5.95 GHz and $f_r$ = 4.45 GHz. (b) $\Delta f$ at single photon regime versus temperature for $f_r$ = 5.95 GHz and $f_r$ = 4.45 GHz.}
 \label{fig_3}
 \end{figure}
We extracted CPW properties from $|S_{21}|$, such as $Q_{i}$ and resonance frequency, ($f_r$), based on the notch-type resonator circuit model \cite{probst2015efficient}. Extracted data confirms that at the single photon regime, $Q_{i}$ initially increases as the temperature rises since TLS defects are primarily responsible for losses in the $T < T_C/10$ regime \cite{foshat2025characterizing}. With further temperature increase, TLS losses become negligible compared to quasiparticle losses, leading to a reduction in $Q_{i}$ and a noticeable shift in $f_r$. Fig.4(a) and Fig.4(b) illustrate the relationship between $Q_{i}$
and $\Delta f$ in the 0.1 K $< T <$ 3 K ranges for both fundamental frequencies. In Fig.4(a), $Q_{i}$ increases from $10^5$ at $T$ = 120 mK to $2.571\times 10^5$ at $T$ = 1 K, and then decreases to $7.421 \times 10^3$ at $T$ = 2.9 K at $f_r$ = 5.95 GHz. Furthermore, Fig.4(b) shows a red shift in the $f_r$ starting between 1.6 K and 1.8 K, which we speculate is due to an increase in the quasiparticle density in the inductive channel.
To evaluate it, we calculate the complex conductivity of the NbN film in the next section.
\subsection{Complex Conductivity Calculation of CPW}
As mentioned earlier, the propagation properties of the resonator, such as $Q_i$ and $f_r$, depend on the CPW's impedance, which is extracted from the conductivity value of the superconductor film. 
  \begin{figure}[!t]
 \centering
%\subfloat[] {\input{Resistance-fr1.pgf}
\hspace{-1 cm}\subfloat[] {\includegraphics[]{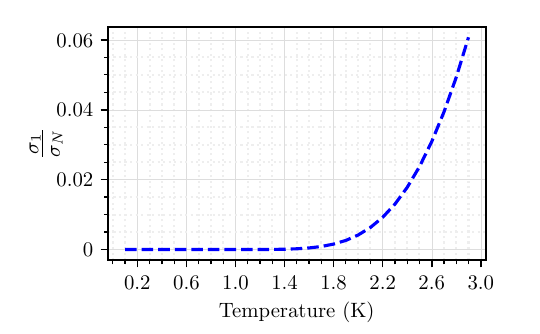}
 \label{fig4_a}}
 
 \vfill
  \hspace{-1 cm}\centering
 %\subfloat[] {\input{Inductance-fr1.pgf}
  \subfloat[] {\includegraphics[]{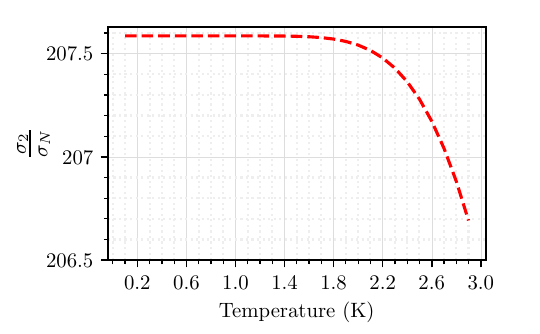}
   \label{fig_4_b}}
 \caption{(a) The real part of the normalized complex conductivity, $\sigma_1$, as a function of temperature. (b) The imaginary part of the normalized complex conductivity, $\sigma_2$, as a function of temperature for the NbN superconducting CPW resonator. Both plots are calculated at $f_r$ = 5.95 GHz.
 }
 \label{fig_4}
 \end{figure}
Superconductor conductivity can be described by the two-fluid model (based on the Mattis-Bardeen theory \cite{mattis1958theory}), where the real part of the conductivity describes the loss in the superconductor film, and the imaginary part exhibits the kinetic inductance value, leading to a resonance frequency shift. The superconductor's complex conductivity, $\sigma (T) = \sigma_{1} (T) - j\sigma_{2}(T)$, when $\hbar\omega \ll \Delta_{0}$ and $k_{B} T \ll \Delta_{0}$, is approximated as \cite{gao2008physics,alexander2021measuring}:
\begin{equation}\label{equation3}
    \frac{\sigma_{1}(T)}{\sigma_N}=\frac{4\Delta_0}{\hbar\omega}e^{(-\frac{\Delta_0}{k_BT})}\text{sinh}(\frac{\hbar\omega}{2k_BT})K_0(\frac{\hbar\omega}{2k_BT})
\end{equation}
\begin{equation*}
    \frac{\sigma_{2}(T)}{\sigma_N}=\frac{4\Delta_0}{\hbar\omega}[1-\sqrt{\frac{2\pi k_B T}{\Delta_0}}\exp{\frac{-\Delta_0}{k_B T}}
\end{equation*}
\par
\begin{equation}\label{equation4}
  -2\exp{\frac{-\Delta_0}{k_B T}}\exp{\frac{-\hbar\omega}{2 k_B T}}I_0(\frac{\hbar\omega}{2 k_B T})]
   \end{equation}

 Where $\sigma_N$ is normal state conductivity, $\Delta_0 = 1.76 k_B T_c$ is the superconducting energy gap at zero temperature, $k_B$ is Boltzman constant, $\hbar$ is reduced plank constant, and $K_0 (x)$ and $I_0(x)$ are modified Bessel functions of second and first order, respectively. 
 \par
 The above equations allow us to calculate the NbN complex conductivity as a function of temperature. Fig.5(a) and Fig.5(b) illustrate $ \sigma_1$ and $ \sigma_2$ in the measured temperature range, respectively.  
Fig.5(a) illustrates the relationship between increasing temperatures and NbN's resistive channel value, $\sigma_1$, which affects the quality factor of CPWs. Moreover, Fig.5(b) describes an inductive channel, $\sigma_2$, of the NbN CPW, influencing the resonator frequency shift. This model is used to describe the complex conductivity of superconducting film and has provided information about the transition from TLS-dominated losses to quasiparticle losses \cite{zmuidzinas2012superconducting}. However, Mattis-Bardeen theory assumes a thermal equilibrium quasiparticle distribution and does not account for non-equilibrium effects such as quasiparticle trapping, diffusion, or photon-assisted generation, which can impact the response of high kinetic inductance materials like NbN.\par
Recent theoretical work has provided analytical descriptions of non-equilibrium quasiparticle distributions in superconducting resonators, revealing that photon-assisted quasiparticle heating and redistribution of quasiparticles can modify $Q_i$ and $f_r$ \cite{fischer2023nonequilibrium}. These effects become particularly relevant in NbN, where high kinetic inductance and increased photon absorption can increase non-equilibrium quasiparticle lifetimes, leading to deviations from conventional Mattis-Bardeen theory.
A fully theoretical model comprising quasiparticle trapping, diffusion, and recombination effects of NbN would be an important future direction. However, the primary objective of this study is to experimentally characterize the quasiparticle density of NbN under varying temperatures, providing a foundation for future theoretical developments.
\begin{figure}[t]
 \centering
\hspace{-0.5 cm}
\subfloat[]
{\includegraphics{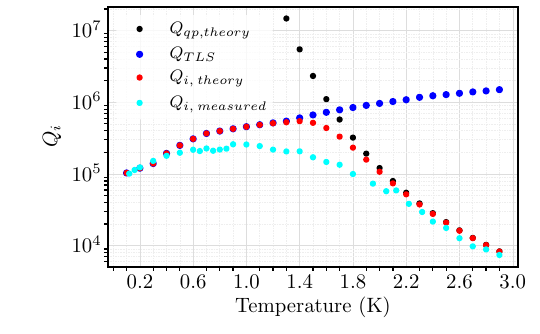}
%\subfloat[] {\input{Figures/conductivity_1_scatter.pgf}
 \label{fig_5_a}}
 \vfill
 \hspace{-0.5 cm}
  \centering
 \subfloat[] 
 {\includegraphics[]{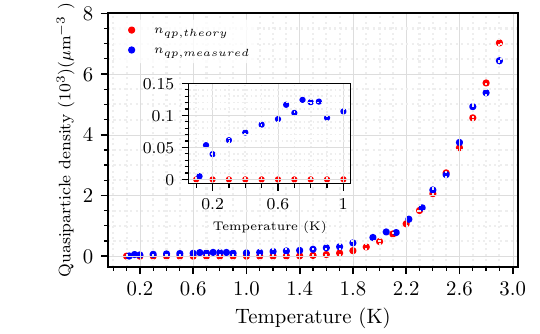}
 % \subfloat[] {\input{Figures/conductivity_2_scatter.pgf}
   \label{fig_5_b}}
 \caption{(a) $Q_{i, measured}$ and $Q_{i, theory}$ versus temperature at $< n_{ph}>$ $\sim$ 1 fitted with the theoretical model of TLS and quasiparticle loss. Both plots are calculated and measured at $f_r$ = 5.95 GHz. (b) Theoretical and measured quasiparticle density of NbN CPW. The inset shows quasiparticle density at $T < T_c/10$.}
 \label{fig_5}
 \end{figure}

\subsection{Thermal and Non-equilibrium Quasiparticle Density}

To demonstrate the existence of non-equilibrium quasiparticle density, we need to compare the theoretical $Q_i$ with the data shown in Fig.4(a). In this section and Fig.6(a), we will differentiate between the measured and theoretically calculated internal quality factors, referring to them as $Q_{i, measured}$ and $Q_{i, theory}$, respectively.
We will estimate $Q_{i, theory}$ within the temperature ranges between 0.1 K $< T <$ 3 K by calculating the TLS loss, $Q_{TLS}$, as evaluated in our previous study \cite{foshat2025characterizing}, as well as the thermal quasiparticle loss from Eq.2
 \begin{equation}\label{equation5}
 \frac {1}{Q_{i, theory} (T)}  = \frac{1}{Q_{TLS}(T)}+\frac{1}{Q_{qp, theory} (T)}
 \end{equation}

Where $Q_{qp, Theroy}(T)^{-1} = \delta_{qp, Theroy}(T)$ is thermal quasiparticle loss coming from Eq.2. Fig.6(a) shows $Q_{qp, Theroy}$, $Q_{TLS}$, $Q_{i, Theroy}(T)$ coming from Eq.2 and $Q_{i, measured}$ adapting from Fig.4(a). In Fig.6(a), there is a deviation between the $Q_{i, theory}$ and $Q_{i, measured}$ in $T < T_c/10$. This deviation suggests the presence of another loss channel coming from non-equilibrium quasiparticle density.\par
To elaborate further, we calculated the non-equilibrium quasiparticle density in the CPWs in the next step. We used the following equations to determine the non-equilibrium quasiparticle density based on the measurement results \cite{barends2011minimizing}:
\begin{equation}\label{equation6}
 \delta_{qp, measured} (T) = \frac{1}{Q_{i, measured}(T)}-\frac{1}{Q_{TLS} (T)}
 \end{equation}
 \par
 \begin{equation*}
    n_{qp, measured}(T) \sim   \delta_{qp, measured} (T) N_0 \Delta (T)
\end{equation*}
 \begin{equation}\label{equation7}
     \frac{\pi}{\alpha} \sqrt{\frac{\hbar \omega}{2 \Delta (T) }}
 \end{equation}
Where $n_{qp, measured}(T)$ is quasiparticle density calculated from measurement results, $N_0 \sim 1.86 \times 10^{28}$ ($\frac{state} {m^{3} eV}$) is the density states at the Fermi level from \cite{chockalingam2008superconducting,hazra2016superconducting}, and $\alpha$ is the kinetic inductance ratio extracted from our previous study \cite{foshat2025characterizing}. 
Additionally, $\delta_{qp, theory} (T)$ in Eq.2 can be utilized to extract quasiparticle density based on Mattis-Bardeen theory with the following equation:
 \begin{equation}\label{equation5}
    n_{qp, theory}(T) \sim \delta_{qp, theory} (T) N_0 \Delta  (T) \frac{\pi}{\alpha} \sqrt{\frac{\hbar \omega}{2 \Delta  (T) }}
\end{equation}

Fig.6(b) illustrates the theoretical and experimental quasiparticle density in 0.1 K $ < T < $ 3 K.
The inset in Fig.6(b) experimentally confirms the presence of non-equilibrium quasiparticle density at temperatures much lower than $T_c$. In literature, quasiparticle density in the superconducting film at $T < T_c/10$ is negligible \cite{wilson2004quasiparticle} because Cooper pair breaking and dissipating heat due to thermal quasiparticles rarely happen; however, we observed noticeable quasiparticle density at millikelvin temperature. In the inset of Fig.6(b), quasiparticle density is saturating at 50 $\mu m^{-3}$ at $T$ = 120 mK, confirming a source of decoherence in quantum circuits at millikelvin temperature. We conclude that the noticeable saturation of quasiparticle density leads to a lower internal quality factor at low temperatures. As shown in Fig. 6(a), $Q_{i, measured}$ is lower than its predicted value. The same behaviour has also been shown in other type of superconducting materials such as Al and TiN superconducting CPWs \cite{alexander2022power,budoyo2016effects,de2011number}. However, NbN exhibits distinct non-equilibrium quasiparticle dynamics due to its shorter electron-phonon relaxation time ($\tau_{e-ph}$) and higher kinetic inductance compared to TiN and Al \cite{riste2013millisecond,guruswamy2014quasiparticle,zmuidzinas2012superconducting}. These properties make NbN particularly sensitive to quasiparticle-induced losses and impedance shifts, distinguishing it from conventional superconductors. Fig.6(a) and Fig.6(b) display the noticeable non-equilibrium quasiparticle impact on superconducting film properties, particularly at millikelvin temperatures, where transmon qubits are measured. This study provides an experimental investigation into these effects, contributing to a deeper understanding of loss mechanisms in high kinetic inductance superconductors. Future works are needed to distinguish between intrinsic saturation and external non-equilibrium contributions.\par
Our microwave loss measurements provide evidence for quasiparticle saturation in superconducting films at millikelvin. However, to fully validate this phenomenon, complementary techniques such as tunneling spectroscopy are critical \cite{joshi2023quasiparticle,fischer2023nonequilibrium}. These methods would allow for more direct measurements of quasiparticle dynamics, helping to distinguish between intrinsic quasiparticle effects and external perturbations coming from cryogenic setups and stray losses. In particular, understanding whether saturation is fundamentally driven by material properties or influenced by environmental factors, like stray loss, is crucial for improving the performance of superconducting circuits. \par

In closing, we want to note that non-equilibrium quasiparticle density can be generated by the interaction between electrons and phonons, resulting in the generation and recombination of quasiparticles. Indeed, microwave readout signals from VNA, essential for device operation, can generate non-equilibrium quasiparticles through photon absorption \cite{barends2011minimizing,de2011number,alexander2021measuring,de2014evidence,morozov2021superconducting}. Note that our study could further extend to analyzing the quasiparticle density in NbN films fabricated via atomic layer deposition. 
\par

%To sum up, understanding thermal and non-equilibrium quasiparticle dynamics is required to better design and engineer unconventional NbN-based superconducting quantum circuits, which the above results help us achieve. 

\section{Conclusion}

In this study, we performed detailed temperature-dependent cryogenic microwave spectroscopy of NbN CPWs resonators operating in single photon regime. Our findings indicate that both TLS and quasiparticle defects impact the performance of NbN superconducting resonators. Additionally, we calculated the complex conductivity of NbN films, capturing the dissipative and dispersive effects on the CPW due to quasiparticle defects.
We then presented $Q_i$ and $f_r$ in response to temperature changes. Subsequently, we verified the effect of thermal and non-equilibrium quasiparticle density on $Q_i$. Our results show that a noticeable non-equilibrium quasiparticle density exists at millikelvin temperatures, leading to decoherence in quantum circuits. 
This research offers insights into the role of quasiparticle dynamics, which is also crucial for advancing quantum error correction techniques in both conventional and unconventional quantum circuits. 
%\vfill

{\appendix[Photon number calculation]
 We can derive the photon number inside the resonator with the below equations \cite{bruno2015reducing}:
\begin{equation}
P_{in} = P_{trans} + P_{reflection} + P_{loss}
\end{equation}

\begin{equation}
P_{loss} = P_{in} (1 - |S_{21}|^2 - |S_{11}|^2)
\end{equation}

\begin{equation}
|S_{21}| = |Q_{c}|^{-2} (|Q_{c}|^2 +Q_{l}^2 - 2 Q_l |Q_{c}|)
\end{equation}

\begin{equation}
|S_{11}| = Q_l^2/|Q_c|^2
\end{equation}

\begin{equation}
< n_{ph} > = Q_i \times \frac{P_{loss}}{\hbar \omega^2}
\end{equation}
\par
Where $< n_{ph} >$  is the photon number, $P_{in} = P_{VNA} + P_{att}$ is the input power inside the chip, $P_{trans}$ is the transmitted power through the feedline, $P_{loss}$ is the source of losses in the chip, $Q_l$ is loaded quality factor, $Q_c$ is coupling quality factor.}
\input{main.bbl}

\end{document}

%% file: main.bbl
% Generated by IEEEtran.bst, version: 1.14 (2015/08/26)